# Nanostructure of a cold drawn tempered martensitic steel

X. Sauvage, X. Quelennec, J.J. Malandain, P. Pareige

Groupe de Physique des Matériaux - UMR CNRS 6634, Institut of Material Research, Université de Rouen, 76801 Saint-Etienne-du-Rouvray, France.

**Abstract**

The carbon atom distribution in a tempered martensitic steel processed by cold drawing was investigated with a three-dimensional atom probe. Data clearly show that cementite starts to decompose at the early stage of deformation. This indicates that the driving force of cementite decomposition during plastic deformation is not related to a strong increase of the interfacial energy. Carbon atmospheres were also analysed. They probably result from pipe diffusion of carbon atoms along dislocations pined by $Fe_3C$ carbides.

## 1. Introduction

Cold drawn pearlitic steel wires are among the strongest commercial steels, and they commonly exhibit a yield stress higher than 3 GPa [1]. Such a high strength is usually attributed to their unique nanostructure : pearlite colonies are strongly elongated along the wire axis during the drawing process and the interlamellar spacing is reduced down to about 20 nm [1-4]. Additionally, it is now well accepted that cementite is also partly decomposed after drawing. The stability of this carbide was first investigated by Mössbauer spectroscopy [5] and in the late 90's this feature was clearly revealed by 3D atom probe analyses [6-11]. The driving force and the kinetics of this phase transformation are still controversial. Some authors point out the possible effect of dislocations which are supposed to trap carbon atoms in cementite lamellae and then to redistribute them in the ferrite phase [5, 13-15]. Others argue that the driving force for the decomposition could be the dramatic increase of the $Fe_3C/\alpha$-Fe interfacial area leading to a strong increase of the carbon solubility in the ferrite through the well known Gibbs-Thomson effect [3, 11]. The main difficulty for the interpretation of experimental data is related to the random orientation of pearlite colonies prior to the drawing process. Depending on the orientation of lamellae with the drawing axis, the deformation of both phases $\alpha$-Fe and $Fe_3C$ is more or less pronounced which gives rise to a wide range of interlamellar spacing in the drawn wire [2, 4] and to the experimentally observed heterogeneous decomposition of cementite [7, 9]. Thus, to clarify the mechanisms of cementite decomposition, this paper reports on experimental data collected in a cold drawn tempered martensitic steel with an homogenous distribution of the $Fe_3C$ phase prior to deformation.

## 2. Experimental

The investigated material is an eutectoid steel with the following composition (at.%): 3.2C-0.37Si-0.32Mn-Fe (balance). Rods were austenized at 1323K, quenched by high pressure air jet and shortly tempered at 823K. This material was then cold drawn up to a true strain of $\varepsilon=3.6$. The resulting microstructure was investigated before deformation and for three different level of plastic deformation ($\varepsilon=0.5, 1.6, 3.6$).

Scanning Electron Microscopy (SEM) specimens were mechanically polished and nital etched (5%HNO₃ in ethanol) before observation in secondary electron mode with a LEO FE 1530 operating at 15kV. Transmission electron microscopy (TEM) specimens were prepared in the cross section of the wires by ion milling. Observations were performed with a JEOL 2000FX microscope operating at 200 kV. 3D atom probe (3D-AP) and field ion microscopy (FIM) specimens were prepared by electropolishing [10]. FIM images and 3D-AP analysis were carried out with a CAMECA's energy compensated optical tomographic atom probe (ECoTAP) in standard conditions [10].

## 3. Results

*Microstructure before drawing ($\varepsilon=0$)*

The microstructure of the material before cold drawing is a typical tempered martensite structure exhibiting both homogenous precipitation of carbides and heterogeneous precipitation on lath boundaries (Fig. 1). Carbides within the laths are roughly spherical shaped with an average diameter of about 50nm, while carbides along lath boundaries are elongated with an aspect ratio of about 5 (length in the range of 100 to 300nm).

Such carbides were analysed with the 3D-AP before drawing and the measured concentrations were always close to 25 at.% as expected for the $Fe_3C$ stoechiometric phase (table 1). In the ferrite phase, the average carbon concentration is four time higher than the solubility of carbon at the tempering temperature (773K) estimated by Thermocalc™ software (table 1). At this temperature, the atomic mobility of Si and Mn atoms is low and during the short tempering treatment the partitioning of these alloying elements is not achieved (table 1). This feature could explain why the carbon concentration in the ferrite is slightly higher than expected.

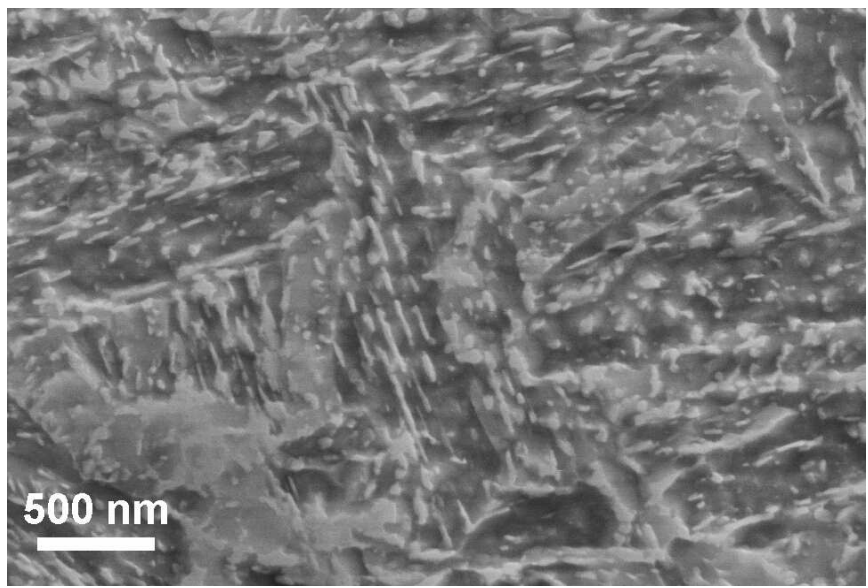

*Figure 1 : SEM image of the microstructure of the tempered martensite before deformation. Cementite particles are brightly imaged.*

|  | at.% C | at.% Mn | at.%Si |
|---|---|---|---|
| Ferrite | | | |
|     Thermocalc™ | 0.01 | 0.07 | 0.41 |
|     3D-AP | 0.04 (± 0.01) | 0.30 (± 0.04) | 0.50 (± 0.05) |
| Cementite | | | |
|     Thermocalc™ | 25 | 2.4 | 0 |
|     3D-AP | 25.1 (± 0.7) | 0.3 (± 0.1) | 0.2 (± 0.1) |

*Table 1 : C, Mn and Si concentration in the ferrite and cementite phases measured by 3D-AP and estimations given by Thermocalc™ software at the tempering temperature (773K).*

## True strain $\varepsilon=0.5$

A cementite particle analysed in the material cold drawn up to a true strain of $\varepsilon=0.5$ is arrowed in the Fig. 2(a) (arrow 1). The carbon concentration profile computed across the $Fe_3C/\alpha$-Fe interface (Fig. 2(b)) shows that the carbon concentration is still about 25at% in the cementite and very low in the ferrite, but a 5 nm wide carbon gradient is exhibited at the interface. Part of this gradient could be attributed to experimental artefacts (thickness of the sampling volume and possible ion trajectory aberrations), but it is obviously much larger than before deformation (see small plot set in Fig. 2(b)). Thus one may conclude that decomposition of cementite has begun and carbon has diffused in the ferrite.

In the Fig. 2(a), a carbon atmosphere is also arrowed close to the carbide (arrow 2). The size of this atmosphere is about 5 nm and its carbon concentration is up to 5 at.% (Fig. 2(c)). As reported by Wilde and co-authors, such features could be typical of a carbon atmosphere around a dislocation [16]. Thus, this could be carbon atoms removed from the $Fe_3C$ particle during the drawing deformation and trapped by a dislocation.

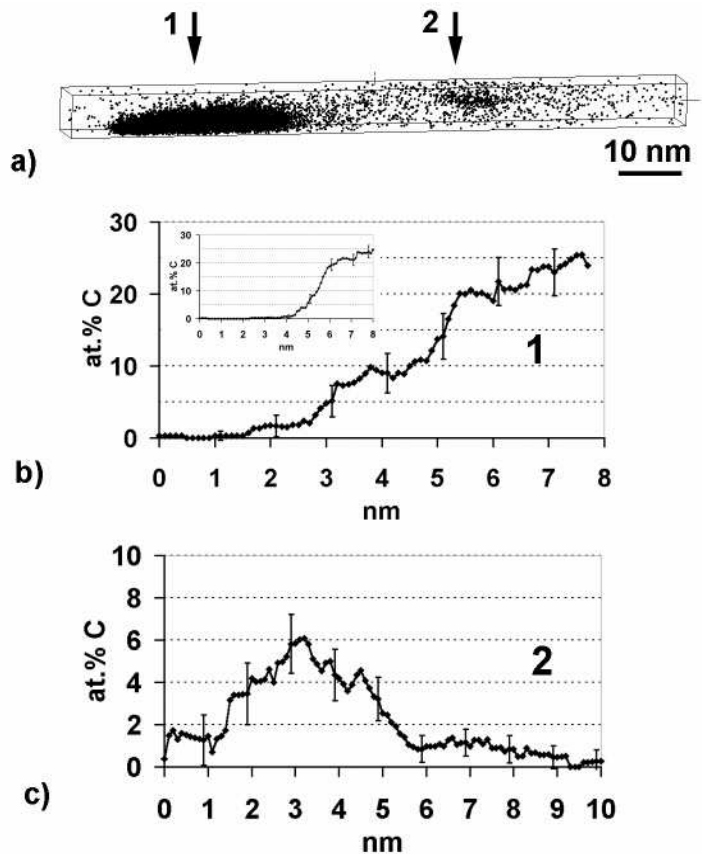

*Figure 2 : (a) 3D reconstructed volume analysed in the material cold drawn up to a true strain of $\varepsilon = 0.5$. Only events associated to C atoms are displayed (black dots) to exhibit a $Fe_3C$ particle (arrow 1) and a carbon rich zone (arrow 2). (b) Carbon concentration profile across the $Fe_3C/\alpha$-Fe interface (arrow 1, Fig.2 (a), thickness of the sampling volume is 1 nm). Inset, carbon concentration profile across such an interface before deformation, given for comparison. (c) Carbon concentration profile across the carbon rich zone (arrow 2, Fig.2 (a), thickness of the sampling volume is 1 nm).*

## True strain $\varepsilon=1.6$

The diffraction pattern in the Fig. 3 clearly shows the Debeye-Scherrer ferrite rings indicating that after drawing up to a true strain of $\varepsilon=1.6$, ferrite grains are in the submicron range and exhibit large mis-orientations. The TEM bright field image reveals elongated ferrite domains (Fig. 3) which are most probably the tracks of martensite lathes. As shown by the (021) $Fe_3C$ ring in the diffraction pattern (arrowhead), there are still many cementite particles. Unfortunately, these particles were not successfully imaged in dark field, probably because they are nanoscaled and because a small aperture had to be used to avoid any overlap with the (110) $\alpha$-Fe ring.

The 3D-AP data set of the Fig. 4 exhibits a lamellae shaped carbon rich region (Fig. 4(a)). Its thickness is in a range of 5 to 10 nm, and it is not possible to estimate its length since it is longer than the analysed volume (75nm). The concentration profile computed across this region indicates that the maximum concentration is close to only 13 at.% (Fig. 4(c)) and a large carbon gradient is also exhibited at the interface. This feature is very similar to $Fe_3C$ lamellae in cold drawn pearlitic steel [7, 9, 11]. However, even if carbon atoms seem to be homogeneously distributed in this lamellae, the data set was filtered to point out possible $Fe_3C$ domains located inside. The filtering procedure could be described as follow : around each atom of the analysed volume, the carbon concentration is measured in a spherical volume (2 nm in diameter). If the value is lower than 23 at.%, the atom is removed from the volume (see reference [16] for details). The obtained 3D map (Fig. 4(b)) shows a large number of nanometer scaled stoechiometric $Fe_3C$ particles within the carbon rich lamellae. Concentration fluctuations related to these particles are not revealed by the profile (Fig. 4(c)) since they are much smaller than the sampling volume which size is optimised to minimised statiscal fluctuations.

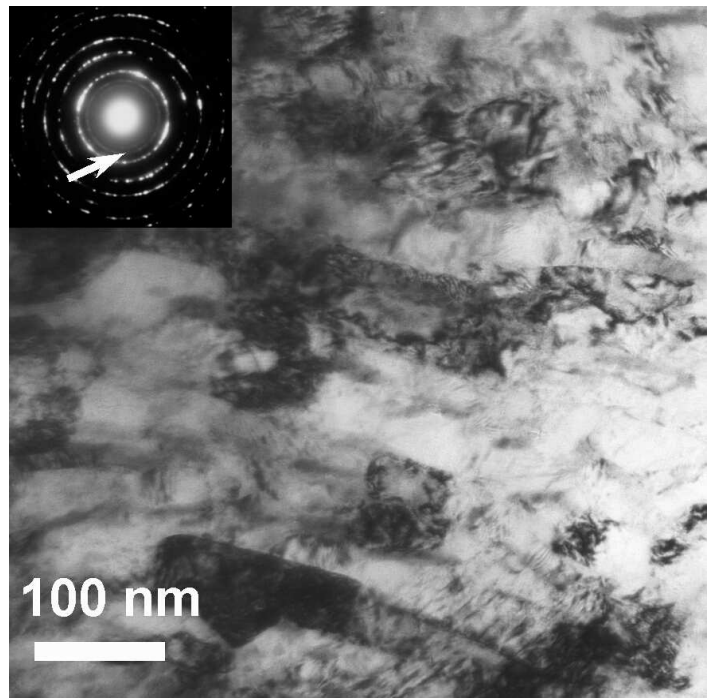

*Figure 3 : TEM bright field image of the cross sectional microstructure after drawing up to a true strain of ε = 1.6. Inset, SAED pattern (aperture 2μm) showing Debeye-Scherrer ferrite rings and (021) $Fe_3C$ ring (arrowhead).*

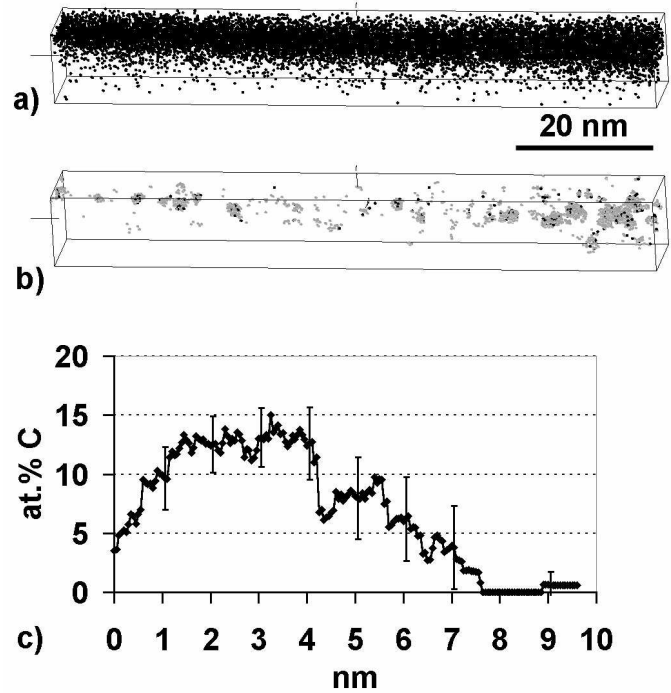

*Figure 4 : (a) 3D reconstructed volume analysed in the material cold drawn up to a true strain of ε = 1.6. Only events associated to carbon atoms are displayed (black dots) to exhibit a carbon rich lamellae. (b) Same 3D volume after filtering : Iron (grey dots) and carbon (black dots) atoms are displayed only where the local concentration is higher than 23 at.% C (see text for details). (c) Carbon concentration profile across the carbon rich lamellae (thickness of the sampling volume is 1 nm).*

*True strain ε=3.6*

At the maximum deformation rate, FIM images of the drawn material exhibit a nanoscaled lamellar structure aligned along the wire axis which is surprisingly very similar to the structure of cold drawn pearlite (Fig. 5(a)). The interlamellar spacing is in the range of 10 to 20 nm and dark lamellae could be $Fe_3C$ or carbon rich zones with a low evaporation field [6-9]. The 3D-AP data set in the Fig. 5(b) shows indeed that they contain a high amount of carbon but the profile across these carbon rich lamellae reveals a concentration lower than after to a true strain of ε=1.6 (about 10at.% instead of about 13at.%). Moreover, no $Fe_3C$ particles were detected thanks to the data filtering described in the previous section. Thus, cementite seems to be completely decomposed at such a high strain but a lamellar nanoscaled structure remains.

In the field ion microscope, an anticlockwise spiral of atomic terraces was imaged on a (110) pole of the α-Fe phase (Fig. 6(a)). This feature is typical of the screw component of a dislocation [16]. The 3D-AP analysis performed along this dislocation clearly exhibits a strong segregation of carbon atoms along this structural defect. The average concentration within the small 3D volume is indeed 2.7 ± 0.1at.%. Thus, as proposed by several authors [5-18-20], dislocations seem to play an important role in cementite decomposition and carbon redistribution in the ferrite phase. This feature would be discussed in the following part.

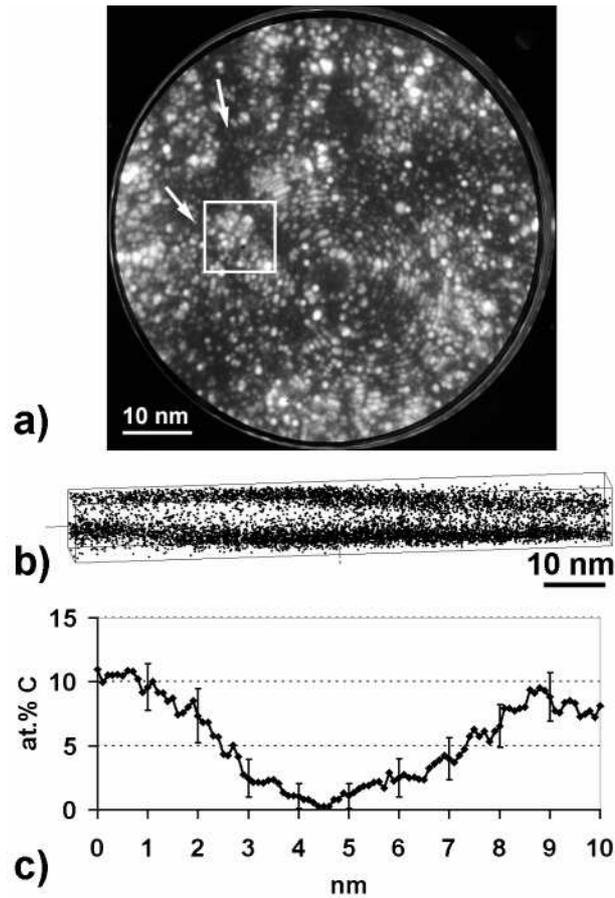

*Figure 5 : (a) FIM picture of the nanostructure after cold drawing up to a true strain of 3.6. Two darkly imaged lamellae are arrowed. (b) 3D-AP data set analysed in the white square displayed in the Fig. 5(a). Only carbon atoms are plotted to exhibit two carbon rich lamellae. (c) Carbon concentration profile across the lamellae (thickness of the sampling volume is 1 nm).*

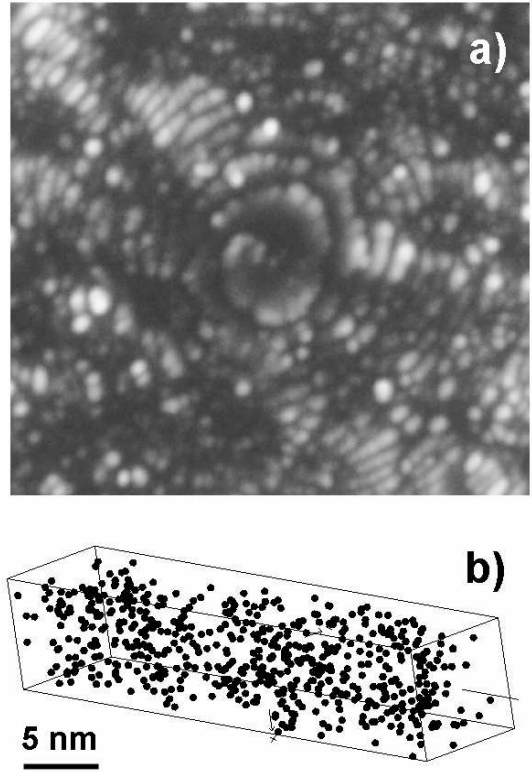

*Figure 6 : (a) FIM picture of a (110) pole of the BCC ferrite phase. The anticlockwise spiral of atomic terraces is characteristic of the screw component of a dislocation. (b) 3D reconstructed volume showing the distribution of carbon atoms around the dislocation (average concentration is 2.7 ± 0.1 at.%).*

4. Discussion

*Progressive decomposition of cementite*

The 3D-AP results have shown that the decomposition of cementite starts at the early stage of the plastic deformation. Indeed, at a true strain of only $\varepsilon = 0.5$, a 5nm carbon gradients was exhibited across a $Fe_3C/\alpha$-Fe interface (Fig. 2). This is almost three times larger than before deformation. This significant increase could be partly attributed to deformation induced interface roughness. However, care was taken to compute this profile in a small area of the interface (3 x 3 $nm^2$) so that this roughness cannot significantly affects this profile. This gradient is therefore attributed to the beginning of cementite decomposition. Inside the carbide, the carbon concentration is still close to 25at.% as expected for the stoechiometric phase $Fe_3C$ (Fig. 2). Unfortunately, 3D-AP data cannot provide any information about the crystallographic structure of the "interface layer" which could be supersaturated $\alpha$-Fe or non stoechiometric Fe$_3$C. Anyway, this layer might contain a significant amount of structural defects like dislocations or vacancies. Fe$_3$C/α-Fe interfaces could be sources and sinks of dislocations, but they are as well strongly modified during the cold drawing process because of the deformation induced texture, the alignment and the curling of ferrite grains along the wire axis. Thus, as reported by Ashby or Gil Sevillano, geometrically necessary dislocations are created along interfaces [17, 18].

At higher level of deformation, carbides are elongated along the wire axis (Fig. 4(a)). The resulting increase of the interfacial area could explain why the carbon gradient does not extend (Fig. 4(c)). Due to the fragmentation of cementite in nanoscaled particles (Fig. 4(b)), it is however impossible to check the influence of the true strain on the slope of this gradient. It is interesting to note that such fragmentation of cementite lamellae has been already reported in cold drawn pearlitic steel [3, 9]. Since interfaces seem to play an important role in the decomposition process, this fragmentation could promote the penetration of the so-called "interface layer" across the carbide and it finally embeds the remaining Fe$_3$C nanoscaled particles (Fig. 4(b)).

At the higher level of deformation, cementite is completely decomposed and the former "interface layer" containing a high amount of carbon on structural defects remains. This gives rise to a lamellar structure very similar to the nanostructure of drawn pearlitic steels (Fig. 5).

Since the decomposition of cementite starts at low level of deformation where carbides are still coarse, the driving force of this reaction cannot be related to a dramatic increase of the surface to volume ratio of this phase (Gibbs-Thomson effect). In fact, it is more likely to be the result of strong modifications of Fe$_3$C/α-Fe interfaces during deformation and the local accumulation of structural defects in a 3 nm wide "interface layer". Since a similar mechanism could lead to cementite decomposition in drawn pearlitic steels, the model published a few years ago by the present authors is therefore probably not realistic [11].

*The role of dislocations*

The mechanism of cementite decomposition assisted by dislocation drag in cold drawn pearlitic steel has been proposed and discussed by Gridnev and Gavrilyuk [5, 18-20]. They argue that the binding enthalpy between carbon atoms and dislocations in ferrite is higher than that between carbon and iron atoms in cementite. So dislocations located close to Fe$_3$C/α-Fe interfaces would trap carbon atoms from Fe$_3$C. These authors assume that binding enthalpies are not affected by internal stresses and high dislocation densities, which has to be proven.

Moreover, the crystallographic transformation following the carbon depletion of the orthorhombic $Fe_3C$ structure into the BCC structure of the ferrite as to be described. Anyway such a mechanism could lead to the formation of carbon gradients analysed with the 3D-AP and discussed in the previous section. However, as shown by the experimental data of the present investigation, dislocations could play an additional role in the case of tempered martensitic steel. At low and high level of deformation, carbon atmospheres around dislocations were indeed revealed by 3D-AP data (Fig. 2(c) and Fig. 6). These dislocations might be pinned by carbides and pipe diffusion of carbon as occurred along the dislocation line. It is well known that the driving force of solute segregation on dislocation core is the strain field resulting from the edge component of dislocations [19]. Thus the two analysed atmospheres exhibit different carbon concentration (respectively about 5 and 3 at.% carbon) because they are probably related to different kind of dislocations.

## 5. Conclusions

i) The decomposition of cementite starts at the early stage of the drawing process (true strain of 0.5) indicating that the driving force for this transformation is not related to the increase of the interfacial energy. ii) At intermediate deformations (true strain of 1.6) only few $Fe_3C$ fragments embedded in carbon rich zones remain. iii) After cold drawing up to a true strain of 3.6, the tempered martensitic steel exhibits a nanostructure very similar to cold drawn pearlitic steels. $Fe_3C$ carbides seem to be completely decomposed, however carbon atoms are not homogeneously distributed. iv) Carbon atmospheres around dislocations were clearly exhibited. Pipe diffusion along dislocations pined by $Fe_3C$ carbides is though to play an important role in the decomposition process.